\documentclass[runningheads]{llncs}
\usepackage{graphicx}
\usepackage{multirow}
\usepackage{array}
\usepackage{svg}
\usepackage{subfigure}

\begin{document}
\title{Studying the Effects of Sex-related Differences on Brain Age Prediction using brain MR Imaging}
% Fairness and Bias in Machine Learning applied to Brain Magnetic Resonance Imaging  \thanks{Supported by organization NSERC.}}
%
\titlerunning{Sex Effects on Brain Age Prediction}
% \author{XXX\inst{1}\orcidID{0000-1111-2222-3333}}

\author{Mahsa Dibaji\inst{1}\orcidID{0009-0004-3166-7737} \and
Neha Gianchandani \inst{2}\orcidID{0000-0003-0822-4554} \and
Akhil Nair\inst{3}\orcidID{0009-0008-9905-969X} \and
Mansi Singhal \inst{4}\orcidID{0009-0009-2395-2956} \and
Roberto Souza\inst{1}\orcidID{0000-0001-7824-5217} \and
Mariana Bento\inst{2}\orcidID{0000-0001-5125-0294}}
%index{Dibaji, Mahsa}
%index{Gianchandani, Neha}
%index{Nair, Akhil}
%index{Singhal, Mansi}
%index{Souza, Roberto}
%index{Bento, Mariana}
%
\authorrunning{M. Dibaji et al.}

\institute{
Department of Electrical and Software Engineering, University of Calgary, Calgary, Alberta, Canada
\and
Department of Biomedical Engineering, University of Calgary, Calgary, Alberta, Canada
\and
Department of Aerospace Engineering, Indian Institute of Technology Kharagpur, Kharagpur, India
\and
Department of Electrical Engineering, Dayalbagh Educational Institute (Deemed University), Agra, India
}

\maketitle              % typeset the header of the contribution
\begin{abstract}

While utilizing machine learning models, one of the most crucial aspects is how bias and fairness affect model outcomes for diverse demographics. This becomes especially relevant in the context of machine learning for medical imaging applications as these models are increasingly being used for diagnosis and treatment planning.

In this paper, we study biases related to sex when developing a machine learning model based on brain magnetic resonance images (MRI). We investigate the effects of sex by performing brain age prediction considering different experimental designs: model trained using only female subjects, only male subjects and a balanced dataset. We also perform evaluation on multiple  MRI datasets (Calgary-Campinas(CC359) and CamCAN) to assess the generalization capability of the proposed models.

We found disparities in the performance of brain age prediction models when trained on distinct sex subgroups and datasets, in both final predictions and decision making (assessed using interpretability models). Our results demonstrated variations in model generalizability across sex-specific subgroups, suggesting potential biases in models trained on unbalanced datasets. This underlines the critical role of careful experimental design in generating fair and reliable outcomes.

%First keyword  \and Second keyword \and Another keyword.}
\end{abstract}
\section{Introduction}

Machine Learning (ML) has shown great promise in healthcare applications, from assisting with diagnoses to informing treatment strategies. However, when deploying ML models in such critical areas, we need to ensure that the algorithms are reliable and do not perpetuate existing biases \cite{chen2023algorithmic}. If an ML model is systematically biased towards a specific demographic group, it can lead to unfair outcomes, and in the worst cases, even harmful consequences \cite{rajkomar2018ensuring}. Such biases could arise from the development data, either due to existing historical biases or the under-representation of certain groups. Even using unbiased data, the outcome might be unfair if protected features are used for prediction. Recognizing and minimizing these biases is essential for creating fair and trustworthy ML models, particularly in healthcare where stakes are high \cite{suresh2021framework,mehrabi2021survey}.

Machine learning's (ML) increasing application in predicting brain age through T1-weighted MRI scans has become a key area of focus \cite{peng2021accurate}, related to brain development, cognitive decline, and neurodegenerative diseases \cite{franke2019ten}. ML models often aim for a `global' brain age index, reflecting brain maturity and serving as a biomarker to assess structural changes and aging \cite{cole2017predicting,wang2019gray}. Exploring what these models learn and the significant brain regions they identify may offer insights into individual brain variations. .

A crucial aspect to consider when developing these models is the sex differences in brain volumes, which could significantly influence their predictions. Existing literature has demonstrated considerable sex-related differences in the total and regional brain volumes, including Gray Matter, White Matter, and Cerebrospinal Fluid \cite{gur1999sex,wang2019effects}. These differences, and their interaction with age, can have significant impacts on cognitive impairment, especially in the elderly \cite{Jurgen14}. Therefore, ensuring our ML models account for these variations and perform equally well across different sex subgroups is crucial for fairness and reliability.

In this paper, we aim to investigate how the performance of brain age prediction models varies across sex subgroups: males only, females only, and balanced datasets. Our goal is to develop accurate and interpretable ML models, while also ensuring their fairness. By understanding how predictions vary across these different subgroups, we seek to offer valuable insights for the development of more reliable and fair ML models, and to promote transparency in ML processes. While focusing on brain age prediction using MRI, our experimental design also can be translatable for other tasks, examining sex differences and dataset variations, enhancing transparency and reliability.

Our experimental design followed a similar approach presented in \cite{larrazabal2020gender}. We trained and validated different models to perform brain age prediction, considering the following experimental design:
a) using only female samples, b) using only male samples, and c) using a balanced mix of both sexes. We also performed experiments using different datasets, with varying data distributions, evaluating inconsistencies for different populations or equity-deserving groups.

% %EXPLAIN PAPER ORGANIZATION AND SECTIONS AND SUBSECTIONS
The rest of the paper unfolds as follows: Section 2 covers materials and methodologies. Section 3 outlines our experimental results. Section 4 discusses these findings and their implications. The paper concludes with a summary and potential future work.

\section{Materials and Methods}
\subsection{Brain MR Datasets}

In this study, we utilized the Calgary-Campinas-359 (CC359) \cite{souza2018open} and the Cambridge Centre for Ageing and Neuroscience (CamCAN) \cite{shafto2014cambridge,taylor2017cambridge} datasets. CC359 is a multi-vendor (General Electric (GE), Philips, and Siemens), multi-field strength (1.5 T and 3 T magnetic field strengths) volumetric brain MRI dataset, comprising 359 T1-weighted three-dimensional (3D) volumes. It has balanced sex distribution, with 183 (50.97\%) female and 176 (49.03\%) male healthy subjects, aged 29 to 80 years. Brain masks are also available for this dataset.

\begin{figure*}
    % \vspace{-6 mm}
    \centering
    \subfigure[]
    {
        \includegraphics[scale=0.5]{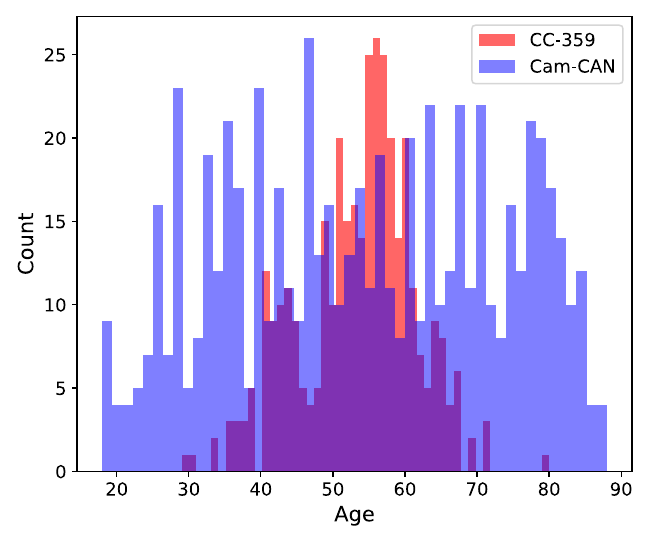}
    }
    \subfigure[]
    {
        \includegraphics[scale=0.5]{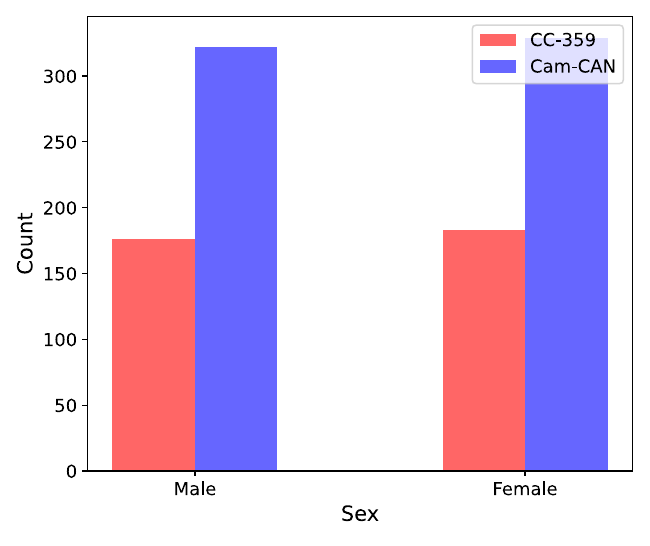}
    }
    %\caption{(a) age distribution (b) sex distribution}
    \caption{Age and sex distribution across the CC359 and CamCAN datasets: (a) The CC359 dataset exhibits an age distribution close to normal and a smaller age range compared to CamCAN. (b) Sex distribution in both datasets reveals a balanced representation of male and female samples.}
    \label{fig:dist}
    % \vspace{-6 mm}
\end{figure*}

% CamCAN description
The CamCAN data set comprises MR images that were collected at a single site (MRC-CBSU) using a Siemens TIM Trio scanner at 3 T magnetic field strength. The dataset is composed of T1-weighted reconstructed brain MR volumes and segmentation masks for certain structures. From the total number of 651 Samples 329 (50.54\%) correspond to female subjects, and 322 (49.46\%) correspond to male subjects aged 18 to 88 years. Fig. \ref{fig:dist} depicts the distribution of age and sex subgroups in CC359 and CamCAN datasets.

\subsection{Pre-processing}
In this study, the preprocessing pipeline consisting of skull stripping, registration, and intensity scaling. We trained a UNet model using the CC359 images and their corresponding brain masks for the skull stripping step. The UNet architecture was chosen for its effectiveness in handling high-resolution brain MR images. This trained model achieved 97\% dice score on the validation set and was then used to perform brain extraction on the CamCAN dataset, thereby effectively stripping non-brain tissues.

We registered the skull-stripped brain MR images to the MNI152 standard atlas \cite{fonov2009unbiased}. This registration was performed using FSL's FLIRT tool \cite{jenkinson2001global,jenkinson2002improved}, which allows for a 6 Degrees-of-Freedom rigid registration. This process involves only rotations and translations without distorting the brain's shape and size. This step ensures that the brain images are comparable and consistent for further analyses. Image intensities were also scaled to fall between 0 and 1.

\subsection{Brain Age Prediction Task}
We utilized a Convolutional Neural Network (CNN) architecture, the Simple Fully Convolutional Network (SFCN) proposed in \cite{peng2021accurate} for estimating brain age based on 3D T1-weighted images. This model comprises seven convolutional blocks. The initial five blocks down-sample the input after each $3\times3\times3$ convolutional layer, followed by a $1\times1\times1$ convolutional block and a classification head. To stabilize the training process, batch normalization is incorporated. The only modification to this model was replacing the classification head with a ReLU-activated linear regressor for brain age prediction.

During the training and validation stages of our study, we employed Mean Absolute Error (MAE) as our loss function. This loss function played a vital role in assessing the disparity between the predicted brain age and the ground truth age labels, evaluating the model's ability to estimate brain age accurately.

\subsection{Grad-CAM Interpretability}
Gradient-weighted Class Activation Mapping (Grad-CAM) \cite{selvaraju2017grad} is an interpretability method for deep learning models, especially CNNs. It highlights important regions in an input image by calculating gradients of the target class score with respect to the final convolutional layer's feature maps. It then creates a heatmap showing regions crucial for the model's decision, offering insights into its behavior. This helps understand why the model makes certain predictions and is useful for transparency and bias detection in critical applications. 

Grad-CAM's interpretability is limited by coarse localization from lower resolution in deeper layers \cite{mcallister2020visualization}. To overcome this limitation, we averaged maps from both early-stage and final convolutional layers. By blending low-level features like textures with high-level insights, this approach offers a  multifaceted interpretation of the model's reasoning, and enhances the robustness of visualizations by potentially mitigating noise. 

\subsection{Experimental Setting} \label{sec:exp}

We established test sets to evaluate the performance of our models through a stratified sampling approach based on vendor and magnetic field, while taking into consideration both sex and dataset. A total of four test sets were constructed in this manner: 30 females from CamCAN ($Cam_F$); 30 males from CamCAN ($Cam_M$); 30 females from CC359 ($CC_F$) and 30 males from CC359 ($CC_M$). 

The remaining samples from each dataset (591 
CamCAN and 299 CC359 samples) were used to create three distinctdevelopment sets: female subjects only, male subjects only, and  larger balanced sets combining both male and female subjects. This resulted in a total of 6 development datasets. Each of these sets was then stratified by vendor and magnetic field strength, and divided into 80\% training and 20\% validation sets. 

\begin{figure}[htbp]

\centerline{\includegraphics[scale=0.8]{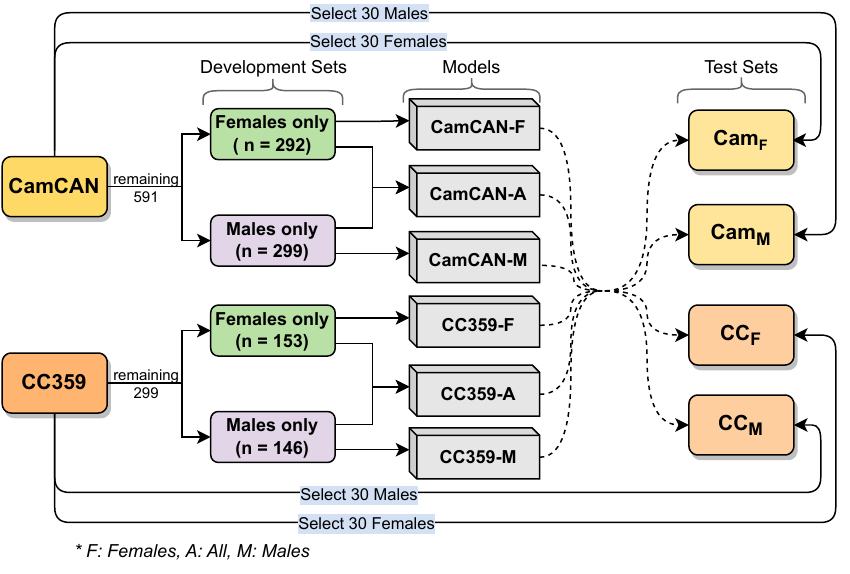}}
\caption{Overview of the experimental design. We utilized 3 development sets: Females only, Males only, and All Subjects, extracted from the CamCAN and CC359 datasets separately. These sets were split in a stratified manner based on vendor and magnetic field strength, resulting in training (80\%) and validation (20\%) subsets. Subsequently, each model was evaluated on four test sets: $Cam_F$, $Cam_M$, $CC_F$, and $CC_M$}
\label{fig:experiment}
\end{figure}

This process led to the generation of six separate models. Three were developed using the CC359 dataset, named as CC359-F (trained on female data), CC359-M (trained on male data), and CC359-A (trained on data from all subjects). The remaining three models were trained on the CamCAN dataset, similarly named as CamCAN-F, CamCAN-M, and CamCAN-A. An illustration of this design is provided in Fig. \ref{fig:experiment}.

To mitigate overfitting, we employed an augmentation step where 50\% of the training and validation samples were randomly subjected to a 15$^{\circ}$ rotation on-the-fly. Training was carried out using the Adam optimizer with a Mean Absolute Error (MAE) loss function and a batch size of 8. The learning rate was initially set to 0.001, then halved every 10 epochs, across a total of 50 epochs.

For our experiments, we implemented data transforms, and Grad-CAM heat maps using MONAI \cite{cardoso2022monai} which is a robust, open-source platform developed by NVIDIA built upon PyTorch. 
The University of Calgary Advanced Research Computing (ARC) cluster, specifically gpu-a100 and gpu-v100, were also utilized.

\section{Results}
Table \ref{tab:cc539-eval} provides a summary of the results (MAE) of models trained on CC359 dataset (CC359-F, CC359-M, and CC359-A), evaluated against the defined test sets: $Cam_F$, $Cam_M$, $CC_F$, and $CC_M$ (detailed in section \ref{sec:exp}). To compare models performances across sex subgroups regardless of dataset source, we created two aggregated test sets, one each for Males ($CC_M$ and $Cam_M$ combined) and Females ($CC_F$ and $Cam_F$ combined). Table \ref{tab:CamCAN-eval} presents similar results for models trained using  CamCAN dataset (CamCAN-F, CamCAN-M, and CamCAN-A). 

\begin{figure}[htbp]
\centerline{\includegraphics[scale=0.7]{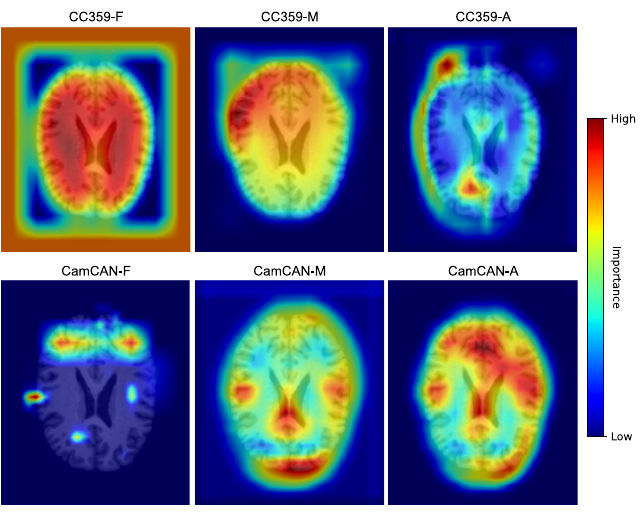}}
\caption{Averaged Grad-CAM heatmaps represents a comparative snapshot of significant regions for brain age prediction between models trained on distinct datasets and sex-specific groups, predicted on Females test set ($CC_F$ and $Cam_F$ combined). Although our model is 3D based, we are showing a 2D slice for visualization and comparison purposes (same slice is shown for each model variation.). Models are identified above each map: CC359-F, CC359-M, CC359-A, CamCAN-F, CamCAN-M, CamCAN-A. }
\label{fig:smap_f}
\end{figure}

In addition, we have produced averaged Grad-CAM maps on each model's predictions for all test sets, using only the aggregated Females test set for visualization and comparisons purposes (Fig. \ref{fig:smap_f}). These maps represent the mean saliency maps across all samples in this test set. They highlight the regions that each model, on average, considered significant for prediction \cite{stanley2022fairness}. The same approach was applied to all test sets, yielding similar outcomes across the board. However, we focused on the Females test set to discuss the observed differences.

\begin{table}[h]
\centering
\caption{Mean Absolute Error (MAE) comparison for three CC359 model: CC359-F (Females Only), CC359-M (Males Only), and CC359-A (All Subjects) across 4 test sets $Cam_F$, $Cam_M$, $CC_F$, $CC_M$, and 2 aggregated sets combining female ($Cam_F$ and $CC_F$) and male ($Cam_M$ and $CC_M$) samples.}
\label{tab:cc539-eval}
\begin{tabular}{l@{\hskip 1cm}c@{\hskip 1cm}c@{\hskip 1cm}c}
\hline
Test Sets & CC359-F & CC359-M & CC359-A \\
\hline
$Cam_F$ & $17.91 \pm 8.864$ & $17.347 \pm 9.387$ & $\mathbf{14.355 \pm 8.117}$ \\
$Cam_M$ & $16.269 \pm 9.452$ & $16.408 \pm 8.983$ & $\mathbf{13.804 \pm 7.564}$ \\
$CC_F$ & $5.588 \pm 5.929$ & $6.47 \pm 5.238$ & $\mathbf{5.428 \pm 5.291}$ \\
$CC_M$ & $8.311 \pm 6.432$ & $7.056 \pm 4.883$ & $\mathbf{5.535 \pm 4.844}$ \\
Males & $12.29 \pm 9.011$ & $11.732 \pm 8.61$ & $\mathbf{9.669 \pm 7.578}$ \\
Females & $11.749 \pm 9.737$ & $11.909 \pm 9.346$ & $\mathbf{9.892 \pm 8.177}$ \\
\hline
\end{tabular}
\end{table}

\begin{table}[h]
\centering
\caption{Mean Absolute Error (MAE) comparison for three CamCAN models: CamCAN-F (Females Only), CamCAN-M (Males Only), and CamCAN-A (All Subjects), across 4 test sets CamF, CamM, CCF, CCM, and 2 aggregated sets combining females (CamF and CCF) and males (CamM and CCM) samples.}
\label{tab:CamCAN-eval}
\begin{tabular}{l@{\hskip 1cm}c@{\hskip 1cm}c@{\hskip 1cm}c}
\hline
Test Sets & CamCAN-F & CamCAN-M & CamCAN-A \\
\hline
$Cam_F$ & $7.801 \pm 6.186$ & $\mathbf{6.105 \pm 4.018}$ & $6.366 \pm 5.386$ \\
$Cam_M$ & $7.002 \pm 4.69$ & $6.504 \pm 4.721$ & $\mathbf{5.763 \pm 4.403}$ \\
$CC_F$ & $9.973 \pm 6.913$ & $9.571 \pm 6.581$ & $\mathbf{9.018 \pm 5.963}$ \\
$CC_M$ & $\mathbf{10.049 \pm 5.949}$ & $12.614 \pm 10.144$ & $10.549 \pm 7.264$ \\
Males & $8.526 \pm 5.569$ & $9.559 \pm 8.481$ & $\mathbf{8.156 \pm 6.466}$ \\
Females & $8.77 \pm 6.63$ & $7.838 \pm 5.721$ & $\mathbf{7.692 \pm 5.834}$ \\
\hline
\end{tabular}
\end{table}

\section{Discussion}

The observed results in Table \ref{tab:cc539-eval}, which relate to CC359 models, and Table \ref{tab:CamCAN-eval}, which relate to CamCAN models, present a considerable degree of variability. Despite being trained for the same task, using the same architecture, and undergoing the same preprocessing within a closely aligned experimental design, divergent outcomes emerged due solely to differences in the development sets.

Our experiments suggest that models trained on a specific dataset fail to generalize effectively to an external dataset. For instance, in Table \ref{tab:cc539-eval}, the MAE of all three models was significantly lower on $CC_F$ and $CC_M$ (sourced from CC359) compared to $Cam_F$ and $Cam_M$. This pattern is also observed on Table \ref{tab:CamCAN-eval}, although the difference in performance is less significant.

The improved generalizability of CamCAN models can likely be attributed to development set size, as shown in Fig. \ref{fig:experiment}. While increased diversity in data (e.g., different vendors and magnetic field strengths) typically improves generalization, this could lead to less accurate predictions if the training samples are insufficient to capture these variations. Therefore, a balance is necessary between dataset size and variability. With increased variability, a larger sample size is required for optimal model function.

The evaluation outcomes on the combined Male and Female test sets demonstrates no consistent trend. Certain models exhibited superior performance on female data, while others were more effective with male data.  Interestingly, these performance variations do not seem to be associated with either the origin of the development data or the sample size.

A closer observation of the model performances on sex-specific test sets (especially females) reveals minimal variation among the F, M, and A model variants, as depicted in each table. To better comprehend whether these models were focused on similar features, we used the GradCAM saliency maps as an interpretability method. Fig. \ref{fig:smap_f} shows that even models trained on a specific dataset identified significantly different features when trained using a single-sex subgroup versus a mixed-sex group. This highlights the importance of model's interpretabiliy to ensure that they are learning the appropriate features.

One model (CC359-F) considered regions outside of the brain highly important (red edges in Fig. \ref{fig:smap_f}), which ideally should not have affected decision-making. For CC359-M, almost all the regions in the brain were relatively important. CC359-A seems to be focused on smaller regions, however, one of those regions is almost outside of the brain. Similarly, CamCAN-F model is also focused on more specific regions and most of the brain tissue is not important in predicting brain age. Interestingly, CamCAN-M seems to be focused on more specific regions compared to CamCAN-A, despite the fact that CamCAN-A had twice as many training samples and was thus anticipated to learn more specific features.
% avoiding 'black-box' models

Our proposal has limitations related to the size of the development sets, potentially impacting the comparisons. However, the performance of models trained on combined males and females was not very different from models trained using only half of the samples (single-sex subgroups). Another limitation is the disregard of age distribution during the split of training, validation, and test sets. Since the CC359 and CamCAN have different age distribution (Fig. \ref{fig:dist}.a), an improved way of splitting training, validation, and test sets should consider the stratified age distribution. Additionally, the use of Grad-CAM, which generates relatively coarse saliency maps may impact the detailed comparison between models and restrict the interpretation of intricate relationships between specific brain regions and predictions. Lastly, the straightforward averaging used to produce an aggregated map for a population presents some challenges, such as potential loss of specificity and alignment complexities. However, these challenges do not diminish the method's promising potential to improve interpretability.

\section{Conclusion}

In this study, we examined the influence of sex and dataset variations on brain age prediction. Our findings emphasize that thoughtful experimental design is crucial in shaping the performance and feature learning of models, leading to outcomes that are both reliable and fair. This underscores the broader need for interpretability methods to ensure trustworthy results. We aimed to evaluate how these variations and sex differences impact the model's performance and generalizability, rather than achieving state-of-the-art accuracy.  We intend to make our code available for easy reproduction and benchmarking of our findings.

Future work should delve into two key areas: a more rigorous examination of performance disparity through statistical tests (e.g., Wilcoxon signed-rank test), complemented by using more precise saliency maps and a more reliable method for aggregating these maps. This will foster greater confidence in conclusions. Concurrently, efforts must be directed towards designing and optimizing predictive models that specifically address sex-related differences. This dual focus aims to reduce biases and ensure reliable and consistent results across varied populations, strengthening the overall impact of the study.

% \paragraph{Sample Heading (Fourth Level)}

% \begin{table}
% \caption{Table captions should be placed above the
% tables.}\label{tab1}
% \begin{tabular}{|l|l|l|}
% \hline
% Heading level &  Example & Font size and style\\
% \hline
% Title (centered) &  {\Large\bfseries Lecture Notes} & 14 point, bold\\
% 1st-level heading &  {\large\bfseries 1 Introduction} & 12 point, bold\\
% 2nd-level heading & {\bfseries 2.1 Printing Area} & 10 point, bold\\
% 3rd-level heading & {\bfseries Run-in Heading in Bold.} Text follows & 10 point, bold\\
% 4th-level heading & {\itshape Lowest Level Heading.} Text follows & 10 point, italic\\
% \hline
% \end{tabular}
% \end{table}

%For citations of references, we prefer the use of square brackets
%and consecutive numbers. Citations using labels or the author/year
%convention are also acceptable. The following bibliography provides
%a sample reference list with entries for journal
%articles~\cite{ref_article1}, an LNCS chapter~\cite{ref_lncs1}, a
%book~\cite{ref_book1}, proceedings without editors~\cite{ref_proc1},
%and a homepage~\cite{ref_url1}. Multiple citations are grouped
%\cite{ref_article1,ref_lncs1,ref_book1},
%\cite{salahuddin2022transparency}.
 \bibliographystyle{splncs04}
 \bibliography{main}
%

%Author, F.: Article title. Journal \textbf{2}(5), 99--110 (2016)

%Author, F., Author, S.: Title of a proceedings paper. In: Editor,
%F., Editor, S. (eds.) CONFERENCE 2016, LNCS, vol. 9999, pp. 1--13.
%Springer, Heidelberg (2016). \doi{10.10007/1234567890}

%Author, F., Author, S., Author, T.: Book title. 2nd edn. Publisher,

%Author, A.-B.: Contribution title. In: 9th International Proceedings
%on Proceedings, pp. 1--2. Publisher, Location (2010)

%LNCS Homepage, \url{http://www.springer.com/lncs}. Last accessed 4
%\end{thebibliography}
\end{document}